\documentclass[prl,twocolumn,superscriptaddress,showpacs,amsmath,amstex,amssymb,citeautoscript]{revtex4-1}
\usepackage{graphicx}
\usepackage{natbib}
\usepackage{amsmath}

\usepackage{amssymb}
\usepackage{appendix}
\usepackage[dvipsnames]{xcolor}
\usepackage[section]{placeins}
\definecolor{darkblue}{rgb}{0,0,.65}
\definecolor{darkgreen}{rgb}{0.28,0.41,0.19}

\usepackage{wrapfig}

\usepackage[%
    pdfauthor={David J. Luitz},%
  pdfstartview=FitH,%
  breaklinks=true,%
  bookmarks=true,%
  colorlinks=true,%
  anchorcolor=black,%
  citecolor=blue,
  filecolor=black,%
  menucolor=black,%
  urlcolor=darkblue,%
  linkcolor=blue,%
 ]{hyperref}
\usepackage[all]{hypcap} %

\begin{document}

\title{An Upper Limit on the Initial Temperature of the Radiation-Dominated Universe}

\author{Betty X. Hu}
\email{bhu@g.harvard.edu}
\affiliation{Department of Physics, Harvard University, 17 Oxford Street, Cambridge, MA 02138, USA}
\author{Abraham Loeb}
\email{aloeb@cfa.harvard.edu}
\affiliation{Department of Astronomy, Harvard University, 60 Garden Street Cambridge, MA 01238, USA}
\date{\today}

\begin{abstract}
Gravitational waves (GWs) are produced by colliding particles through the gravitational analogue of electromagnetic bremsstrahlung. We calculate the contribution of free-free emission in the radiation-dominated Universe to the stochastic GW background. We find that the energy density of the resulting GW radiation is heavily dependent on the number of elementary particles, $N_{\mathrm{tot}}$, and the maximum initial temperature, $T_{\mathrm{max}}$. We rule out $N_{\mathrm{tot}}\gtrsim N_{\mathrm{SM}}$ for $T_{\mathrm{max}}\sim T_{\mathrm{Planck}}\approx10^{19}$ GeV and $N_{\mathrm{tot}}\gtrsim10^{13}\times N_{\mathrm{SM}}$ for $T_{\mathrm{max}}\sim10^{16}$ GeV, where $N_{\mathrm{SM}}$ is the number of particles in the Standard Model. In the case of inflation, existing cosmological data constrain $T_{\mathrm{max}}\lesssim10^{16}$ GeV. However, alternative models to inflation such as bouncing cosmologies allow for $T_{\mathrm{max}}$ near $T_{\mathrm{Planck}}$. At the energy scales we are considering, the extra number of particles arise naturally in models of extra dimensions. 
\end{abstract}
\maketitle

\textbf{Introduction.}
The stochastic gravitational wave (GW) background originates from independent physical processes throughout the history of the Universe \cite{christensen2019}. Potential sources include the coalescence of compact binaries \cite{ligo1, ligo2, ligo3}, quantum fluctuations during inflation \cite{easther2006, easther2007, cook2012}, cosmic strings \cite{siemens2007, depies2007}, first order phase transitions in the early Universe \cite{kosowsky1992, kamionkowski1994}, and other exotic phenomena \cite{buonanno2005, marassi2009}. Ref. \cite{christensen2019} provides a comprehensive review summarizing the sources, current and planned observation methods, and constraints on the stochastic GW background.

The GW background spans a wide range of frequencies. The cosmic microwave background (CMB) anisotropies probe GWs with frequencies $f\sim10^{-20}-10^{-15}$ Hz \cite{lasky2016, planck6, kamionkowski1997}; pulsar timing arrays such as the EPTA, PPTA, and NANOGrav are sensitive at $f\sim10^{-9}-10^{-7}$ Hz \cite{kramer2013, yardley2010, mclaughlin2013}; ground-based detectors such as LIGO and Virgo at $f\sim10^{1}-10^{3}$ Hz \cite{pitkin2011, ligo4, christensen2019}, and space-based detectors such as LISA will be sensitive at $f\sim10^{-4}-1$ Hz \cite{bartolo2016, amaroseoana2013}. Although less often discussed in the literature, there has been growing interest in building detectors sensitive to $f\sim10^8$ Hz \cite{nishizawa2008, cruise2006, akutsu2008} and even higher frequencies \cite{grishchuk2003}. As such, it is instructive to study potential sources of higher-frequency GWs. 

Bremsstrahlung radiation from the collision of charged particles \cite{jackson} is analogous to GWs produced through the gravitational scattering of two particles, a process known as free-free emission. Gravitational free-free emission has been studied classically in the context of, for example, high-speed black hole encounters \cite{death1978}, massive objects \cite{kovacs1978}. More recently, the case of free-free emission from collisions of massless particles has been considered as well\cite{gruzinov2014, spirin2015}. 

Here, we study the contribution of free-free emission in the radiation-dominated Universe to the stochastic GW background. Conservatively, we ignore other sources of high frequency GWs in the range of $10^{10}-10^{15}$ Hz \cite{bisnovatyi_kogan2004}, such as thermal GWs from stars \cite{weinberg}, the amplification of quantum fluctuations of the gravitational field in the early Universe by inflation \cite{allen1997, grishchuk2007}, graviton to photon conversion in the presence of large scale magnetic fields \cite{pshirkov2009}, thermalized photons converting into gravitons in the presence of strong primordial magnetic fields \cite{fujita2020}, primordial black hole evaporation \cite{inomata2020}, and black hole mergers in the early Universe \cite{hooper2020}.

In the case of cosmological inflation \cite{guth1981, sato1981, brout1978}, data from the \textit{Planck} satellite \cite{planck10} and BICEP2/Keck \cite{ade2018} provide constraints on the tensor-to-scalar ratio, $r$, which, together with the amplitude of the power spectrum of the primordial scalar perturbations, $A_s$, can be translated to an upper bound on the energy scale of the inflationary potential when the pivot scale exits the Hubble radius, $V_*<(1.6\times10^{16}\,\mathrm{GeV})^4$. The resultant energy density of a thermal bath of massless particles after reheating can be expressed as $V_*=g_*(T)T^4\pi^2/30$, where $g_*(T)$ is the total number of effectively massless degrees of freedom and $T$ is the temperature of the thermal bath. With $g_*(T_{\mathrm{max}})\approx100$ \cite{kolbturner}, we can constrain the maximum post-reheating temperature of the radiation-dominated Universe, $T_{\mathrm{max}}\lesssim10^{16}\,\mathrm{GeV}$. 

Here, we derive equivalent constraints on $T_{\mathrm{max}}$ for alternatives to inflation. As an example, in bouncing cosmologies the universe began in a contracting phase, experienced a bounce, and eventually entered the radiation-dominated phase of Big Bang cosmology \cite{brandenberger2017}. In these models, $T_{\mathrm{max}}$ during the bounce can be near the Planck scale \cite{brandenberger2020}, $T_{\mathrm{Planck}}\approx10^{19}$ GeV. We assume for simplicity the standard cosmological history of a single matter-dominated phase following the radiation-dominated epoch and ignore more complicated alternatives \cite{chen2006}.

\textbf{Free-free GW emission.}
We define the present-day GW energy density per logarithmic frequency interval relative to the closure density of the universe as
\begin{equation} \label{eq:1}
\ \Omega_{\mathrm{GW}}(f)=\frac{1}{\rho_c}\frac{d\rho_{\mathrm{GW}}}{d\,\mathrm{ln}f},\
\end{equation}
where $\rho_{\mathrm{GW}}$ is the energy density of the radiation and $\rho_c$ the critical density,
\begin{equation} \label{eq:2}
\ \rho_c=\frac{3H_0^2c^2}{8\pi G},\
\end{equation}
where $H_0$ is the Hubble constant today, $G$ Newton's constant, and $c$ the speed of light. Throughout our work, we adopt $h=H_0/(100\,\mathrm{km}\,\mathrm{s}^{-1}\mathrm{Mpc}^{-1})=0.7$ \cite{planck6, riess2018}. 

The observed energy density of the GWs per logarithmic frequency interval is,
\begin{equation*} \label{eq:3}
\ \frac{d\rho_{\mathrm{GW}}}{d\,\mathrm{ln}f}=\int\Gamma\frac{dE_{\mathrm{GW}}}{d\,\mathrm{ln}f}\frac{1}{(1+z)^4}dt\
\end{equation*}
\begin{equation} \label{eq:3}
\ =\int\Gamma\frac{dE_{\mathrm{GW}}}{d\,\mathrm{ln}f}\frac{1}{H(z)(1+z)^5}dz,\
\end{equation}
where $\Gamma$ is the collision frequency per proper volume, $E_{\mathrm{GW}}$ is the GW energy, $t$ is cosmic time, and $z$ is redshift. The factor $(1+z)^{-3}$ normalizes to comoving volume, an additional factor $(1+z)^{-1}$ accounts for the redshifting of the energy \cite{phinney2001}, and $H(z)$ is the Hubble parameter. Transforming $dE_{\mathrm{GW}}/d\,\mathrm{ln}f$ to the source rest frame with $f_r=f(1+z)$,
\begin{equation*} \label{eq:4}
\ \frac{d\rho_{\mathrm{GW}}}{d\,\mathrm{ln}f}=f\int\Gamma\frac{dE_{\mathrm{GW}}}{df}\frac{1}{H(z)(1+z)^5}dz\
\end{equation*}
\begin{equation} \label{eq:4}
\ =f\int\Gamma(1+z)\frac{dE_{\mathrm{GW}}}{df_r}\frac{1}{H(z)(1+z)^5}dz,\
\end{equation}
from which we get,
\begin{equation} \label{eq:5}
\ \Omega_{\mathrm{GW}}(f)=\frac{f}{\rho_c}\int\Gamma\frac{dE_{\mathrm{GW}}}{df_r}\frac{1}{H(z)(1+z)^4}dz.\
\end{equation}
The collision frequency per proper volume is, $\Gamma=n_1n_2\sigma c$,
with $n_1$, $n_2$ being the number densities and $\sigma$ the interaction cross section. For massless gauge-boson exchanges \cite{kolbturner}, 
\begin{equation} \label{eq:6}
\ \sigma\sim\left(\frac{\alpha\hbar c}{k_BT}\right)^2,\
\end{equation}
with $\alpha$ being the gauge coupling constant, $\hbar$ the reduced Planck constant, $k_B$ the Boltzmann constant, and $T$ the temperature of the cosmic plasma. For collisions of photons, the number densities are
\begin{equation} \label{eq:7}
\ n_1=n_2=g\int\frac{d^3p}{\left(2\pi\right)^3}f(p)\approx\frac{g\left(k_BT\right)^3}{\pi^2\left(\hbar c\right)^3},\
\end{equation}
allowing us to write the collision frequency per proper volume as,
\begin{equation} \label{eq:8}
\ \Gamma\approx\left[\frac{g\left(k_BT\right)^3}{\pi^2\left(\hbar c\right)^3}\right]^2\left(\frac{\alpha\hbar c}{k_BT}\right)^2c.\
\end{equation}

Ref. \cite{gruzinov2014} calculated classical gravitational free-free emission from the gravitational scattering of two massless particles at leading order in the center of mass deflection angle $\theta\ll1$. Under the approximation that the characteristic scale of the angular frequency $\omega$ is $c^5/GE$,
\begin{equation} \label{eq:9}
\ \frac{dE_{\mathrm{GW}}}{d\omega_r}=\frac{4}{\pi}\theta^2E^2\frac{G}{c^5}\,\mathrm{ln}\left(\frac{c^5}{E\omega_r G}\right),\
\end{equation}
with $\omega_r=2\pi f_r=2\pi f(1+z)$ being the source frame angular frequency. This result holds in the range $c/b<\omega_r<c^5/GE$, with $b$ the impact parameter. However, energy conservation imposes a stricter upper bound of $\omega_r\lesssim3k_BT/\hbar$, the average total energy of an extreme relativistic gas. We note that the aforementioned assumption that $\omega\sim c^5/GE$ holds only for $T_{\mathrm{max}}\lesssim0.1\,T_{\mathrm{Planck}}$.

The center of mass deflection angle, $\theta$, is a function of the impact parameter $b$. In the relativistic regime, $b_{\mathrm{min}}\approx h/p=hc/E$, and we approximate $b\sim b_{\mathrm{min}}$. The classical deflection angle is $\theta=8GE/bc^4$ \cite{spirin2015}, where $E=k_BT=k_BT_0(1+z)$, with $T_0=2.725$ K being the present-day CMB temperature. With these parameters, we can use Eqs. (\ref{eq:8}) and (\ref{eq:9}) to integrate Eq. (\ref{eq:5}) from present time, $z=0$, to the end of inflation, $z_{\mathrm{max}}\sim T_{\mathrm{max}}/T_0$. This ignores the modest (order unity) heating of the CMB but not of the GW background by annihilations (similar to $e^+e^-$ heating of the CMB relative to the neutrino background).

We only calculated $\Omega_{\mathrm{GW}}(f)$ for the case of photon collisions. However, near the Planck scale we ignore order unity corrections that distinguish other types of particles from photons. Instead, we multiply the expression for $\Omega_{\mathrm{GW}}(f)$ in Eq. (\ref{eq:5}) by the number of elementary particles. Most of the contribution to $\Omega_{\mathrm{GW}}(f)$ comes from $z\sim z_{\mathrm{max}}\sim T_{\mathrm{max}}/T_0$. 

In deriving our results we focus on two ways that the contribution of free-free emission to the stochastic GW background can be enhanced: (i) through the number of elementary particles, which can constrain physics beyond the Standard Model, and (ii) through the proximity of $T_{\mathrm{max}}$ to $T_{\mathrm{Planck}}$. In our results, we express the temperature relative to the Planck scale, $T_{\mathrm{Planck}}=\sqrt{\hbar c^5/Gk_B^2}\approx1.2\times10^{19}$ GeV.

\textbf{Constraints on free-free GW emission.}
Big Bang Nucleosynthesis (BBN) and the CMB can be used to set upper bounds on the total energy density of a cosmological GW background \cite{caprini2018},
\begin{equation} \label{eq:10}
\ \Omega_{\mathrm{GW}}=\int\Omega_{\mathrm{GW}}(f)d\ln f=\frac{\rho_{\mathrm{GW}}}{\rho_c}.\
\end{equation}
The GW background acts as an additional component of the radiation field in the Hubble expansion rate,
\begin{equation} \label{eq:11}
\ H(z)=H_0\left[\left(\Omega_{\mathrm{rad}}+\Omega_{\mathrm{GW}}\right)\left(1+z\right)^4+\Omega_{\mathrm{m}}\left(1+z\right)^3+\Omega_{\Lambda}\right]^{1/2},\
\end{equation}
where $\Omega_{\mathrm{rad}}$, $\Omega_{\mathrm{m}}$, and $\Omega_{\Lambda}$ are contributions of the standard radiation (CMB + neutrino background), matter, and the cosmological constant, respectively. BBN and the CMB probe the cosmic energy budget and constrain $\Omega_{\mathrm{GW}}$. The change in the radiation energy density, $\rho_{\mathrm{rad}}$, is then $\Delta\rho_{\mathrm{rad}}\equiv\left(\rho_{\mathrm{rad}}-\rho_c\Omega_{\mathrm{rad}}\right)$. This can be expressed in terms of $\Delta N_{\nu}$ extra neutrino species,
\begin{equation} \label{eq:12}
\ \Delta\rho_{\mathrm{rad}}=\frac{\pi^2}{30}\frac{7}{4}\Delta N_{\nu}T^4.\
\end{equation}
By requiring $\rho_{\mathrm{GW}}\leq \Delta\rho_{\mathrm{rad}}$, we get,
\begin{equation} \label{eq:13}
\ h^2\Omega_{\mathrm{GW}}\leq5.6\times10^{-6}\Delta N_{\nu}=5.6\times10^{-6}\Delta N_{\mathrm{eff}},\
\end{equation}
with $N_{\mathrm{eff}}$ the effective number of neutrino species present in the thermal bath after $e^+e^-$ annihilation. 

Assuming GWs with homogeneous initial conditions, the \textit{Planck} satellite and other cosmological data constrain $\Omega_{\mathrm{GW}}\lesssim4\times10^{-7}$ \cite{caprini2018}.

\textbf{Results.}
To gauge the contribution to $\Omega_{\mathrm{GW}}$ per logarithmic redshift interval, we examine
\begin{equation*} \label{eq:14}
\ \frac{d\Omega_{\mathrm{GW}}(f)}{d\ln (1+z)}=\frac{d^2\Omega_{\mathrm{GW}}}{d\ln (1+z)d\ln f}\
\end{equation*}
\begin{equation} \label{eq:14}
\ =\frac{f}{\rho_c}\Gamma\frac{dE_{\mathrm{GW}}}{df_r}\frac{1}{H(z)(1+z)^3}.\
\end{equation}

Assuming $H(z)\approx H_0\sqrt{\Omega_{\mathrm{rad}}}(1+z)^2$ in the radiation dominated era and defining the following constants,
\begin{equation*} \label{eq:15}
\ \Gamma_0\equiv\left[\frac{g\left(k_BT_0\right)^3}{\pi^2\left(\hbar c\right)^3}\right]^2\left(\frac{\alpha\hbar c}{k_BT_0}\right)^2c,\
\end{equation*}
\begin{equation*} \label{eq:15}
\ \theta_0\equiv\frac{8G(k_BT_0)^2}{hc^5},\
\end{equation*}
\begin{equation*} \label{eq:15}
\ \varepsilon_0\equiv2\pi\left(\frac{4}{\pi}\right)\theta_0^2(k_BT_0)^2\frac{G}{c^5},\
\end{equation*}
\begin{equation} \label{eq:15}
\ \phi_0(f)\equiv\frac{c^5}{k_BT_02\pi Gf},\
\end{equation}
we can write Eq. (\ref{eq:14}) as
\begin{equation*} \label{eq:17}
\ \frac{d^2\Omega_{\mathrm{GW}}}{d\ln (1+z)d\ln f}=\frac{f\varepsilon_0\Gamma_0}{\rho_cH_0\sqrt{\Omega_{\mathrm{rad}}}}\ln\left[\frac{\phi_0(f)}{(1+z)^2}\right](1+z)^5\
\end{equation*}
\begin{equation} \label{eq:16}
\ \propto\ln\left[\frac{\phi_0(f)}{(1+z)^2}\right](1+z)^5.\
\end{equation}
Numerically, we find that Eq. (\ref{eq:14}) is well-described by a power law,
\begin{equation} \label{eq:17}
\ \frac{d^2\Omega_{\mathrm{GW}}}{d\ln (1+z)d\ln f}\propto z^{\beta},\
\end{equation}
with $\beta\approx4.75$ over the observed frequency range. It is evident that the highest values of $z$ dominate the contribution to $\Omega_{\mathrm{GW}}(f)$. Hence, one can simply examine the dependence of $\Omega_{\mathrm{GW}}$ on $T_{\mathrm{max}}$.

\begin{figure}[h]
	\centering
	\includegraphics[width=\columnwidth]{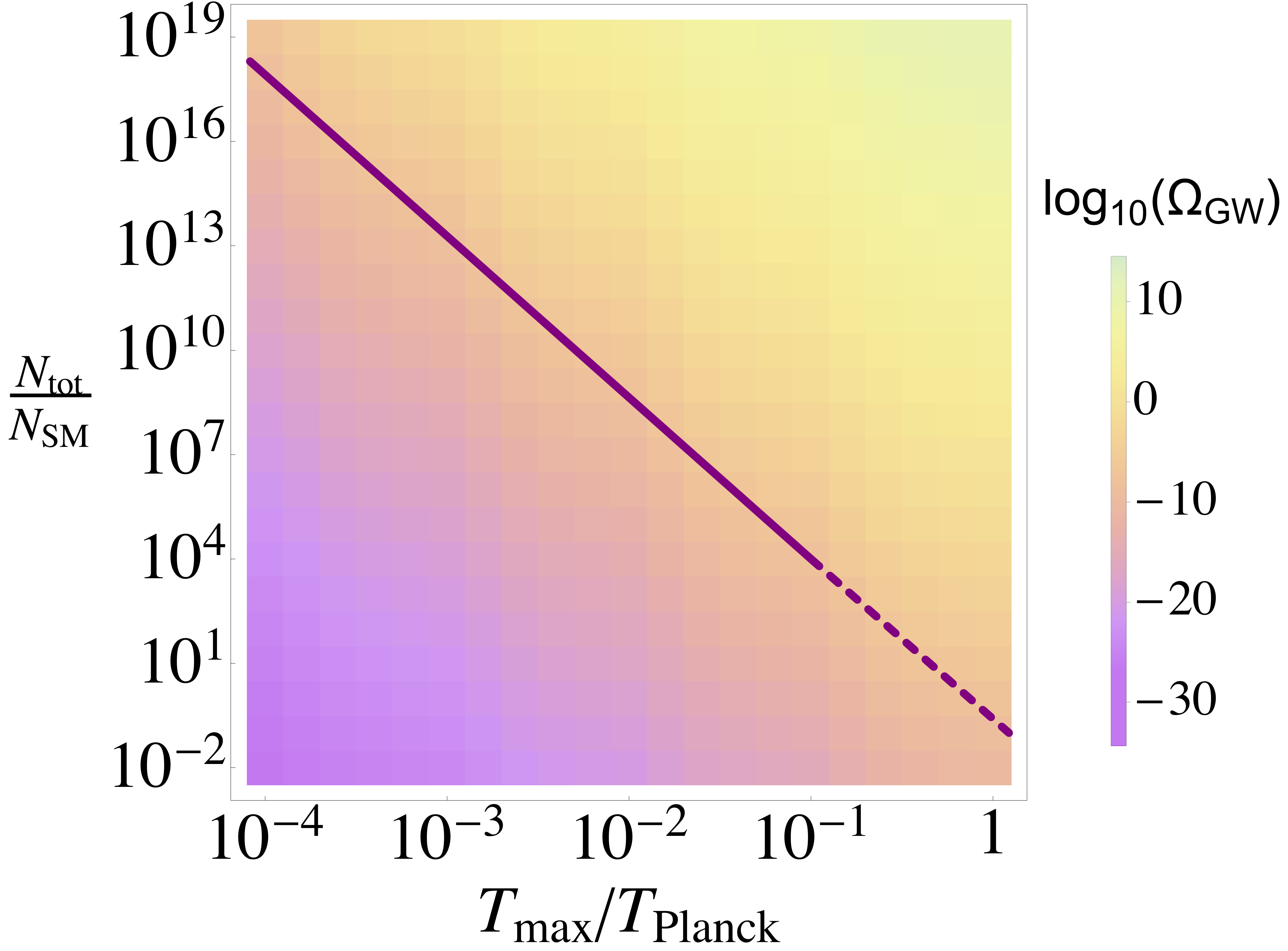}
	\caption{$\log_{10}\Omega_{\mathrm{GW}}$ as a function of $N_{\mathrm{tot}}/N_{\mathrm{SM}}$ and of $T_{\mathrm{max}}/T_{\mathrm{Planck}}$. In general, as either $N_{\mathrm{tot}}$ or $T_{\mathrm{max}}$ increases, the energy density of the radiation increases. The solid purple line indicates the upper bound given by Eq. (\ref{eq:13}), with the region above and to its right ruled out based on existing cosmological data. The dotted line is obtained from extrapolating the solid line, and lies in the region where Eq. (\ref{eq:9}) loses its validity.
	\label{fig:1}}
\end{figure}

Figure \ref{fig:1} shows $\log_{10}\Omega_{\mathrm{GW}}$ as a function of the total number of elementary particles divided by the number of elementary particles in the Standard Model, $N_{\mathrm{tot}}/N_{\mathrm{SM}}$, where $N_{\mathrm{SM}}=17$, and as a function of the maximum temperature of the Universe normalized by the Planck temperature, $T_{\mathrm{max}}/T_{\mathrm{Planck}}$. The solid diagonal line is the upper limit given by Eq. (\ref{eq:13}); the region above and to its right is ruled out from cosmological data. The dashed diagonal line in Fig. \ref{fig:1} is the solid line extrapolated to higher tem- peratures, where the assumption that $\omega\sim c^5/GE$, used in deriving Eq. (\ref{eq:9}), is no longer valid.

As either $N_{\mathrm{tot}}$ or $T_{\mathrm{max}}$ increases, the energy density of the radiation increases. For $T_{\mathrm{max}}$ near $T_{\mathrm{Planck}}\sim10^{19}$ GeV, the cosmological constraint rules out particle physics models that predict $N_{\mathrm{tot}}\sim N_{\mathrm{SM}}$.

\begin{figure}[h]
	\centering
	\includegraphics[width=\columnwidth]{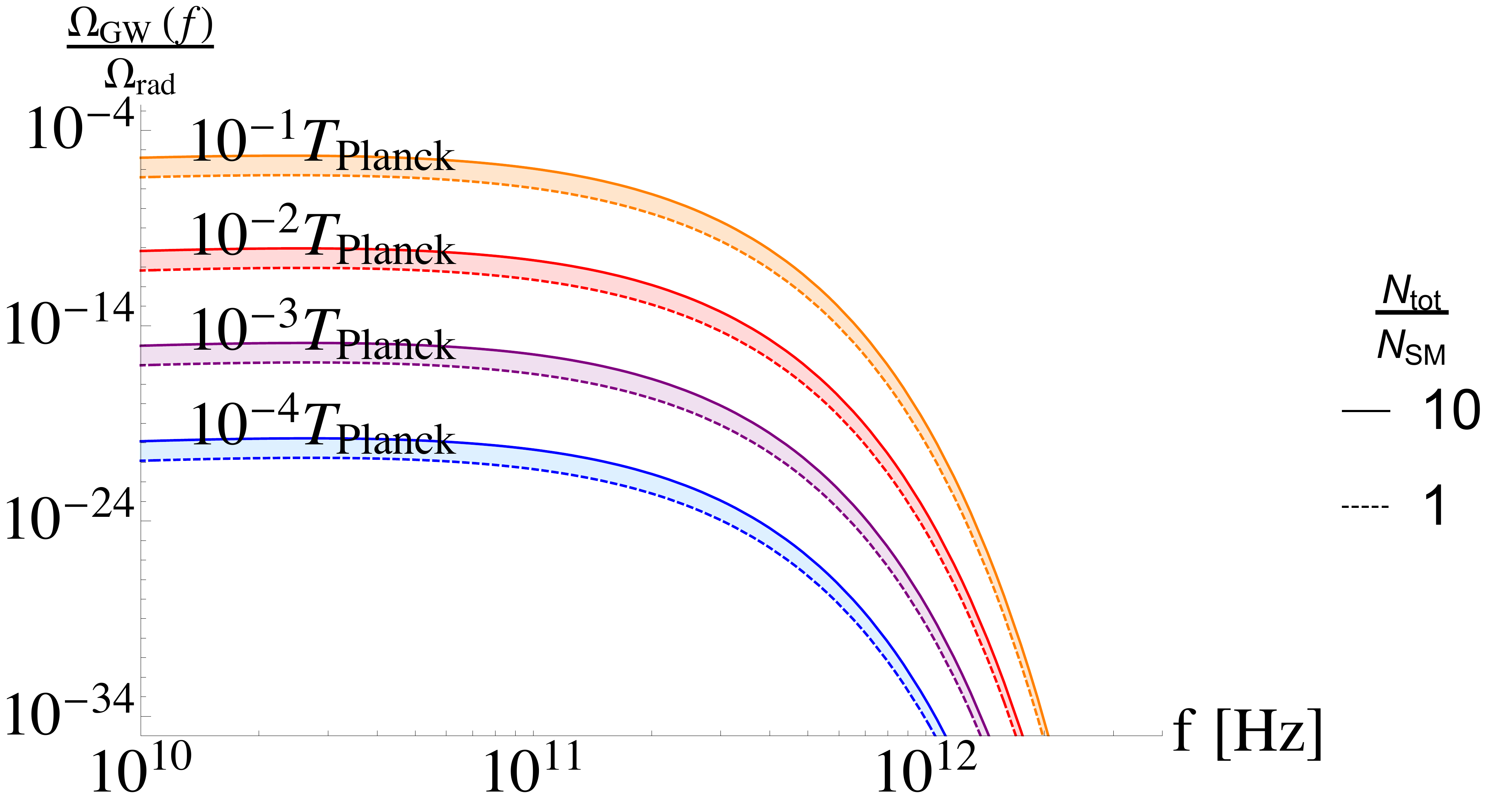}
	\caption{The ratio of GW to the radiation energy density, $\Omega_{\mathrm{GW}}(f)/\Omega_{\mathrm{rad}}$, with varied numbers of elementary particles $N_{\mathrm{tot}}$ and varied maximum initial temperature $T_{\mathrm{max}}$. Each colored band represents a different $T_{\mathrm{max}}$, expressed relative to $T_{\mathrm{Planck}}\sim10^{19}$ GeV. The upper boundary of each band is for $N_{\mathrm{tot}}=10\times N_{\mathrm{SM}}$, and the lower boundary of each band is for $N_{\mathrm{tot}}=N_{\mathrm{SM}}$. The observer frequency is thermally cut off past $f\sim10^{12}$ Hz. 
	\label{fig:2}}
\end{figure}

Figure \ref{fig:2} shows the ratio of GW to the radiation energy density, $\Omega_{\mathrm{GW}}(f)/\Omega_{\mathrm{rad}}$, versus observed GW frequency. Each band in Fig. \ref{fig:2} represents a different value of $T_{\mathrm{max}}$, which we vary from $10^{-4}T_{\mathrm{Planck}}\sim10^{15}$ GeV to $T_{\mathrm{Planck}}\sim10^{19}$ GeV. The lower boundary of each band is for $N_{\mathrm{tot}}=N_{\mathrm{SM}}$ and the upper boundary is for $N_{\mathrm{tot}}=10\times N_{\mathrm{SM}}$.

There is a thermal cutoff near an observed frequency of $f\sim10^{12}$ Hz, which is similar to the frequency cutoff of the CMB. The lower boundary on $f$ originates from the smallest possible value of the impact parameter $b_{\mathrm{min}}$ for classical free-free emission. Over this frequency range, $\Omega_{\mathrm{GW}}(f)$ varies weakly as a function of $f$, but varies strongly as a function of $N_{\mathrm{tot}}$ or $T_{\mathrm{max}}$. 

\textbf{Discussion.} We find that the energy density of the GW background from free-free emission in the early Universe can be used to rule out $N_{\mathrm{tot}}\gtrsim N_{\mathrm{SM}}$ for $T_{\mathrm{max}}\sim10^{19}$ GeV. If we assume a higher-dimensional Planck scale such as $\overline{M}_{4+n}\sim10^{3}$ GeV, near the lower bound imposed from astrophysics and cosmology \cite{hannestad2002, cullen1999, barger1999, hanhart2001}, about $10^{10}$ Kaluza-Klein (KK) states could easily be accessed even for $n$ as low as one \cite{cheng2010, satheesh2006}. From a four-dimensional point of view, these KK excitations are distinct particles, meaning this extra number of particles could easily arise from theories that predict ten, eleven, or more dimensions \cite{stringtheory, dienes1997, maharana2013}. 


However, we caution that our work concerns energy scales at which four-dimensional effective field theories may no longer be accurate \cite{georgi}. An intrinsically higher-dimensional description of free-free GW emission might be necessary. We also note that in theories with a large number $N$ of particle species, it has been pointed out that black hole physics imposes an upper bound on the energy at which these theories are valid, $T\sim M_{\mathrm{Planck}}/\sqrt{N}$ \cite{dvali2008}. The upper bounds we place on $N_{\mathrm{tot}}$ as a function of $T_{\mathrm{max}}$ lie beyond this energy limit for $T_{\mathrm{max}}\lesssim10^{18}$ GeV. 

\textbf{Acknowledgments.}
We thank Xingang Chen and Matt Reece for useful discussions. This work was supported by the Black Hole Initiative at Harvard University, which is funded by grants from the John Templeton Foundation and the Gordon and Betty Moore Foundation.

\bibliography{ref.bib}

\end{document}